\documentclass[10pt,twocolumn]{article}
\usepackage[a4paper, margin=1.25cm, left=1.5cm, right=1cm, bottom=2cm]{geometry}
\usepackage{amsmath, amssymb, amsthm}
\usepackage{booktabs}
\usepackage{float}     
\usepackage{typst}    
\usepackage{xcolor}
\usepackage{listings}
\lstdefinestyle{latex}{alsoletter={\\,_,:,@}, keywords={}, keywordsprefix={\\},
  keywordstyle=\bfseries\color[rgb]{0,0.5,0}, commentstyle=\itshape\color[rgb]{0.24,0.48,0.48},
  morecomment=[l]\%}
\newcommand{\tex}{\lstinline[style=latex]}
\newcommand{\code}{\lstinline}
\usepackage[colorlinks=true, hypertexnames=false]{hyperref}
\usepackage[backend=biber, style=numeric]{biblatex}
\hypersetup{ urlcolor=blue, linkcolor=blue, citecolor=blue }
\newcommand{\repolink}{\href{https://github.com/MrP01/EveryAuthorIsFirstAuthor}{github.com/MrP01/EveryAuthorIsFirstAuthor}}

\makeatletter
\def\input@path{{../lib/}{lib/}}
\makeatother
\usepackage{equalauthors}

\theoremstyle{plain}
\newtheorem{proposition}{Proposition}
\newtheorem{corollary}[proposition]{Corollary}
\theoremstyle{definition}
\newtheorem{definition}[proposition]{Definition}

\newcommand{\equalAuthorsBase}{
  \equalauthorsAddAuthor{Peter Waldert}
  \equalauthorsAddAuthor{Benedikt Kantz}
  \equalauthorsAddAuthor{Thomas K\"ohler}
  \equalauthorsAddAuthor{Erik D. Demaine}
  \equalauthorsAddAuthor{Martin L. Demaine}
}
\newcommand{\equalAuthorsTest}{
  \equalAuthorsBase
  \equalauthorsAddAuthor{Jes\'us Salvador Mart\'inez Alcaide}
}

\equalauthorsSetup{alpha=1, beta=1, smin=0.85, smax=1.15}
\equalauthorsAddAuthor{Peter Waldert}
\equalauthorsAddAuthor{Benedikt Kantz}
\equalauthorsAddAuthor{Thomas Köhler}

\title{Truly Every Author is First Author: Equal-Width Superimposition by Minimising Visual Distortion}
\author{\equalauthorsAuthors}
\date{\today}

\begin{document}
  \makeatletter
  \twocolumn[
    \begin{@twocolumnfalse}
      \maketitle
      \begin{center}
        \begin{minipage}{0.87\textwidth}
          \small
          \begin{abstract}
            In an earlier work, \citeauthor{demaine2023every}~\cite{demaine2023every} propose that every author be a first author by superimposing all author names on the same spot of the title page.
            The construction is typographically unstable, because names have different natural widths.
            The superimposed block is ragged and the longest name dominates the visual mass, reintroducing exactly the ordering advantage the standard removes.
            We close this loophole.
            Each name is distorted by two knobs, a horizontal glyph scale and an inter-letter tracking, until all names occupy one common width $W$.
            We give a quadratic distortion model, solve the per-name trade-off in closed form by Lagrange multipliers, and show that the total-distortion optimal common width is a stiffness-weighted mean of the natural widths.
            We provide implementations as a \LaTeX{} package and a Typst package, covering author blocks and bibliography entries alike; this paper typesets its own author block and its own reference with the former.
            Every proposition is machine-checked in Lean.
            The \LaTeX{} and Typst packages, and the proofs are available at \repolink.
          \end{abstract}
        \end{minipage}
      \end{center}
      \vspace{1.5\baselineskip}
    \end{@twocolumnfalse}
  ]
  \makeatother

  \section{Introduction}
  The convention that author order encodes credit is a lossy encoding of a partial order into a total one.
  \citeauthor{demaine2023every} \cite{demaine2023every} resolve it by making the encoding degenerate, superimposing all author names at the same position, so that no name precedes another.

  AMOR~\cite{weiherer2024amor} takes the opposite route and keeps the list, but shuffles it at random whenever it is viewed, so that no order is fixed.
  This removes the ranking across views but not within any single one, where some name still comes first; superimposition removes it in every view, and is the construction we build on.

  Superimposition removes the horizontal ordering but not the horizontal \emph{extent}.
  If, for example, Yu Liu and Martin L.~Demaine are overprinted, the shorter name is contained in the longer one, and the reader's eye recovers a ranking from ink coverage, where the widest name frames the block and the narrowest hides inside it.
  Formally, superimposition quotients out translation but leaves width as a visible, and thus creditable, per-author quantity.
  In Tufte's terms~\cite{tufte1983visual}, the block has an unbounded lie factor: it displays a width difference where the data, equal credit, has none.

  We therefore require that all names be typeset to one common width.
  Since names cannot be padded with whitespace - padding is invisible under superimposition and merely restores the containment problem - the width must be achieved by distorting the letters themselves.

  Two distortions are available, and they differ in character.
  Scaling the glyphs horizontally changes letterform, while tracking changes only the gaps.
  Both are ugly in excess.
  The engineering question is how to divide a required width change between them, and what common width to aim at in the first place.

  Sections~\ref{sec:model} and~\ref{sec:solution} answer both in closed form.
  Section~\ref{sec:impl} describes the implementations and a machine-checked proof of every claim, and Section~\ref{sec:results} the results.

  \section{Problem Statement}
  \label{sec:model}
  Fix a font and a size.
  All lengths are in printer's points.
  The model is stated for a single name and carries no index; the index appears only where names are compared.

  \begin{definition}[Name data]
    For a name we write
    \begin{itemize}
      \item $g \in \mathbb{N}_{\ge 0}$, the number of inter-character gaps, that is,
            the number of positions at which tracking is inserted;
      \item $c = g + 1$, the number of characters;
      \item $w > 0$, the natural set width, as the font sets it;
      \item $e = w / c$, the mean character width, which is the natural length
            scale of the name.
    \end{itemize}
    Section~\ref{sec:impl} obtains $w$ and $g$ by measurement, without inspecting the characters of the name.
  \end{definition}

  \begin{definition}[Distortion variables]
    A name is typeset with a horizontal glyph scale $s > 0$, under which every character is drawn $s$ times as wide at unchanged height, and a tracking $t \in \mathbb{R}$, an extra kern of length $t$ inserted into each of the $g$ gaps.
    The undistorted setting is $(s,t) = (1,0)$.
  \end{definition}

  Because the scale acts on characters and the tracking on gaps, the two contributions are additive and the realised width of the distorted name is exactly
  \begin{equation}
    \omega(s,t) \;=\; s\,w + g\,t .
    \label{eq:width}
  \end{equation}
  The symbol $\omega$ is reserved for the width a name actually attains, and $W$ below for the common width all names are asked to attain.

  \begin{definition}[Distortion cost]
    With weights $\alpha,\beta > 0$, the ugliness of a name is
    \begin{equation}
      C(s,t) \;=\; \tfrac{\alpha}{2}\,(s - 1)^2 \;+\; \tfrac{\beta}{2}\,\Bigl(\frac{t}{e}\Bigr)^{\!2}.
      \label{eq:cost}
    \end{equation}
  \end{definition}

  Both terms are quadratic and vanish at the undistorted setting, so small distortions are cheap and the cost grows without bound.
  The scale term is dimensionless by construction.
  The tracking term is made dimensionless by measuring $t$ in units of the mean character width $e$: a kern of one average letter is equally offensive in a large font and a small one, and equally offensive in a name of five letters and one of twenty.
  The ratio $\beta/\alpha$ is the only free parameter of the model and expresses the designer's relative tolerance: $\beta/\alpha \to \infty$ forbids tracking and recovers pure scaling, $\beta/\alpha \to 0$ forbids scaling and recovers pure tracking.

  \subsection{Multiple Names}
  A document has $n \ge 1$ names, indexed by $i \in \{1,\dots,n\}$, and from here on the quantities of name $i$ carry that index, so that $g_i$, $c_i$, $w_i$, $e_i$, $s_i$, $t_i$, $\omega_i$ and $C_i$ mean for name $i$ what $g$, $c$, $w$, $e$, $s$, $t$, $\omega$ and $C$ mean for a single name, and the same applies to quantities introduced later, such as the stiffness $K$ of Section~\ref{sec:split}.

  \begin{definition}[Equal-width superimposition problem]
    Given $\{(w_i,g_i,e_i)\}_{i=1}^{n}$ and $\alpha,\beta > 0$, find a common width $W > 0$ and per-name distortions $(s_i,t_i)$ solving
    \begin{equation}
      \begin{aligned}
        \min_{W,\,(s_i,t_i)} \quad & \sum_{i=1}^{n} C_i(s_i,t_i)                  \\
        \text{s.t.} \quad          & \omega_i(s_i,t_i) = W, \quad i = 1,\dots,n .
      \end{aligned}
      \label{eq:problem}
    \end{equation}
  \end{definition}

  Section~\ref{sec:solution} solves \eqref{eq:problem} in two stages, the inner problem for fixed $W$, then the outer problem over $W$.

  \section{Closed-Form Solution}
  \label{sec:solution}
  The problem separates.
  For a fixed common width the per-name trade-off has an explicit minimiser (Section~\ref{sec:split}), whose optimal value is quadratic in that width, so the outer choice of width is an explicit weighted mean (Section~\ref{sec:width}).

  \subsection{The Optimal Split for a Fixed Width}
  \label{sec:split}
  This subsection concerns one name at a time, so it needs no index.
  Fix the target width $W$, and write $x = s - 1$ for the scale excess and $d = W - w$ for the width deficit; the constraint $\omega(s,t) = W$ becomes $w x + g t = d$, while the cost \eqref{eq:cost} reads $\frac{\alpha}{2}x^2 + \frac{\beta}{2}(t/e)^2$ in these variables.

  \begin{definition}[Stiffness]
    The \emph{stiffness} of a name is
    \begin{equation}
      K \;=\; \frac{w^2}{\alpha} \;+\; \frac{g^2 e^2}{\beta} \;>\; 0 .
      \label{eq:stiffness}
    \end{equation}
  \end{definition}

  $K$ measures how cheaply a name absorbs a width change.
  It is large when the name is long, since a given absolute change is then a small relative change, and small for short names.
  This is why $K$ appears as an inverse weight below.

  \begin{proposition}[Per-name split]
    \label{prop:split}
    For fixed $W$, the unique minimiser of \eqref{eq:cost} subject to $\omega(s,t) = W$ is, with $\lambda = (W - w)/K$,
    \begin{equation}
      s \;=\; 1 + \frac{\lambda w}{\alpha}, \qquad t \;=\; \frac{\lambda g e^2}{\beta},
      \label{eq:split}
    \end{equation}
    and the optimal value is
    \begin{equation}
      C^\star(W) \;=\; \frac{(W - w)^2}{2 K}.
      \label{eq:cstar}
    \end{equation}
  \end{proposition}

  \begin{proof}
    The objective is strictly convex and the constraint affine, so the Karush-Kuhn-Tucker conditions are necessary and sufficient.
    With Lagrangian
    \[
      \mathcal{L} = \frac{\alpha}{2}x^2 + \frac{\beta}{2}\frac{t^2}{e^2} - \lambda\,(w x + g t - d),
    \]
    stationarity gives $\alpha x = \lambda w$ and $\beta t / e^2 = \lambda g$, that is, $x = \lambda w/\alpha$ and $t = \lambda g e^2/\beta$.
    Substituting into the constraint,
    \[
      d = w x + g t = \lambda\Bigl(\frac{w^2}{\alpha} + \frac{g^2 e^2}{\beta}\Bigr) = \lambda K ,
    \]
    so $\lambda = d/K$, which is \eqref{eq:split}.
    The value follows from $\frac{\alpha}{2}x^2 + \frac{\beta}{2}(t/e)^2 = \frac{\lambda}{2}(wx + gt) = \frac{\lambda d}{2} = d^2/(2K)$.
  \end{proof}

\begin{figure}[t]
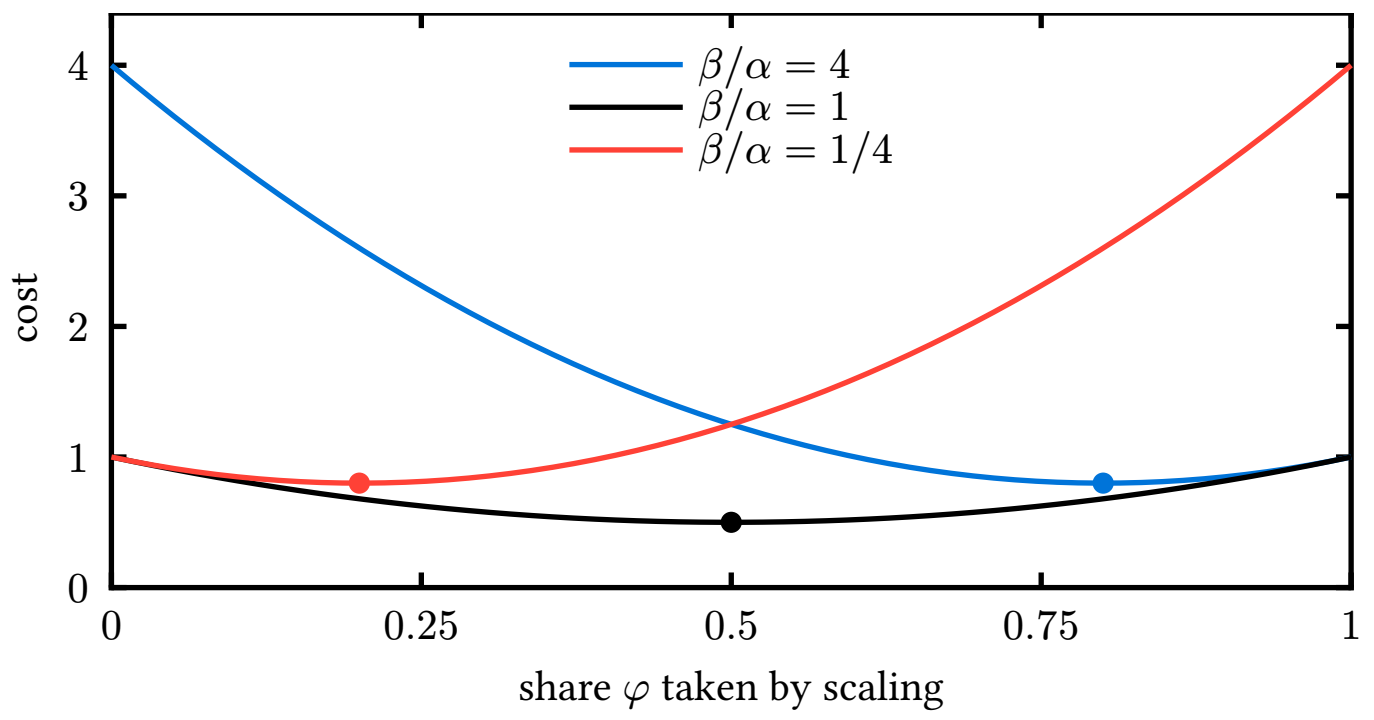

    \centering
    \begin{typst}{width=\columnwidth}
      #import "@preview/cetz:0.4.0"
      #import "@preview/cetz-plot:0.1.2": plot
      #set text(8pt)

      // Cost of splitting a fixed width deficit d between scaling (share phi) and
      // tracking (share 1 - phi), in units of d^2 / (2 w^2), for three weightings.
      #let cost(a, b, phi) = a * phi * phi + b * (1 - phi) * (1 - phi)

      #cetz.canvas({
      import cetz.draw: *
      plot.plot(
      size: (8.2, 3.8),
      x-label: [share $phi$ taken by scaling], y-label: [cost],
      x-min: 0, x-max: 1, y-min: 0, y-max: 4.4,
      x-tick-step: 0.25, y-tick-step: 1, y-minor-tick-step: none,
      legend: "inner-north", legend-style: (stroke: none, fill: white.transparentize(15
      {
          for (a, b, col, lbl) in (
          (1.0, 4.0, blue,  $beta\/alpha = 4$),
          (1.0, 1.0, black, $beta\/alpha = 1$),
          (4.0, 1.0, red,   $beta\/alpha = 1\/4$),
          ) {
              plot.add(domain: (0, 1), samples: 60, style: (stroke: col), label: lbl,
              phi => cost(a, b, phi))
              let ph = b / (a + b)
              plot.add(((ph, cost(a, b, ph)),), mark: "o",
              mark-style: (fill: col, stroke: none), mark-size: 0.14)
            }
        },
      )
      })
    \end{typst}
    \caption{The Cost of Splitting a Fixed Width Deficit.
      The share $\phi$ of the deficit taken by scaling costs
      $\alpha \phi^2 + \beta (1-\phi)^2$ in units of $d^2/2w^2$, and the minimum
      (dot) sits at $\phi^\star = \beta/(\alpha+\beta)$, so the weights set the
      split, up to the approximation $g e \approx w$ of the split ratio \eqref{eq:ratio}.}
    \label{fig:split}
  \end{figure}

  Two consequences are worth stating.
  First, the two distortions are used in the fixed ratio
  \begin{equation}
    \frac{g t}{(s - 1) w} \;=\; \frac{\alpha\, g^2 e^2}{\beta\, w^2} \;\approx\; \frac{\alpha}{\beta} ,
    \label{eq:ratio}
  \end{equation}
  where the approximation uses $g e \approx w$ and is exact up to one character.
  The width change is therefore split between tracking and scaling in essentially the ratio $\alpha : \beta$, independently of $W$, as Figure~\ref{fig:split} shows; the weights are directly interpretable, and $\alpha = \beta = 1$ means ``half the correction each''.
  Second, $C^\star$ is quadratic in $W$, which makes the outer problem trivial.

  \subsection{The Optimal Common Width}
  \label{sec:width}
  \begin{proposition}[Optimal width]
    \label{prop:width}
    The minimiser of $F(W) = \sum_i C_i^\star(W)$ is the stiffness-weighted mean of the natural widths,
    \begin{equation}
      W^\star \;=\; \frac{\sum_{i=1}^{n} w_i / K_i}{\sum_{i=1}^{n} 1 / K_i},
      \label{eq:wstar}
    \end{equation}
    and it is unique.
  \end{proposition}

  \begin{proof}
    By \eqref{eq:cstar}, $F(W) = \frac12 \sum_i (W - w_i)^2 / K_i$ is a strictly convex quadratic in $W$ with $F''(W) = \sum_i K_i^{-1} > 0$.
    Setting $F'(W) = \sum_i (W - w_i)/K_i = 0$ and solving for $W$ gives the weighted mean \eqref{eq:wstar}.
  \end{proof}

  \begin{corollary}
    $\min_i w_i \le W^\star \le \max_i w_i$, with equality only if all $w_i$ coincide.
  \end{corollary}

\begin{figure}[t]
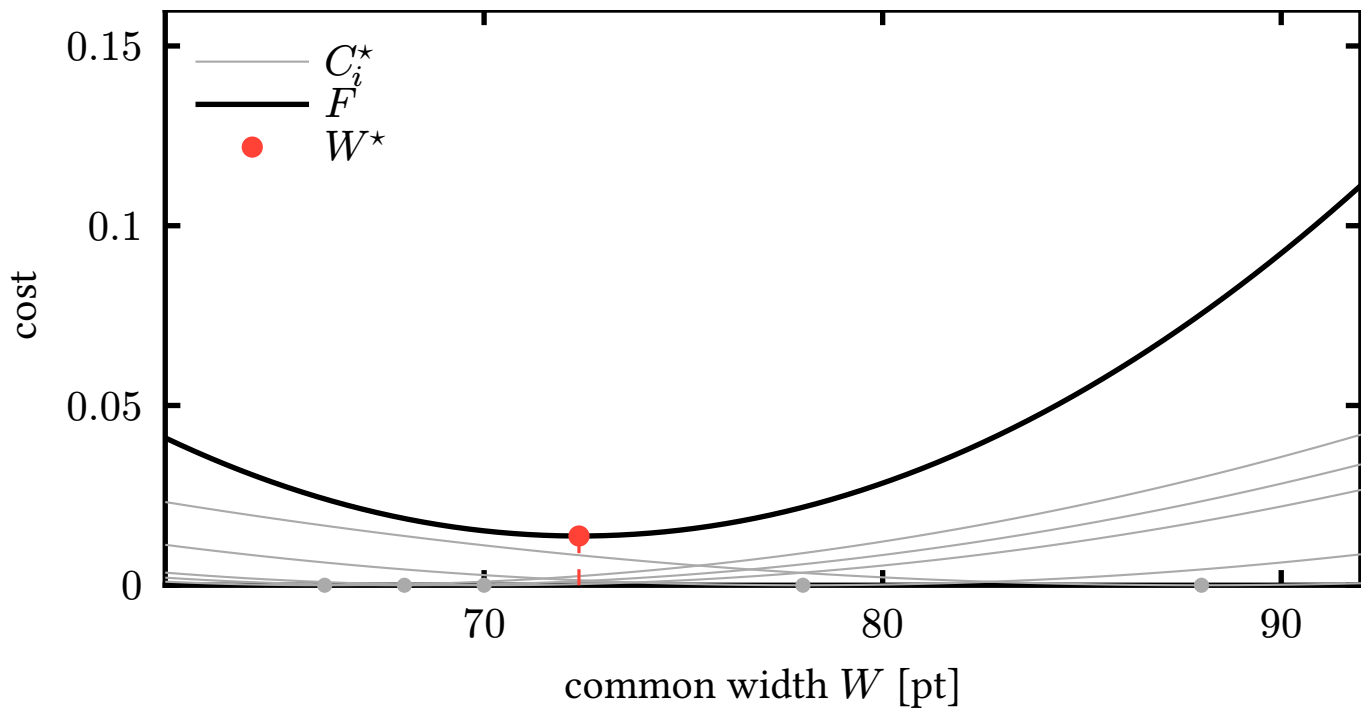

    \centering
    \begin{typst}{width=\columnwidth}
      #import "@preview/cetz:0.4.0"
      #import "@preview/cetz-plot:0.1.2": plot
      #set text(8pt)

      // Five names, given by natural width w and character count c.
      #let names = ((66.0, 13), (70.0, 14), (68.0, 13), (78.0, 15), (88.0, 17))
      #let alpha = 1.0
      #let beta = 1.0
      #let stiff(w, c) = {
      let e = w / c
      let g = c - 1
      w * w / alpha + g * g * e * e / beta
      }
      #let ks = names.map(((w, c)) => stiff(w, c))
      #let ws = names.map(((w, c)) => w)
      #let wstar = ws.zip(ks).map(((w, k)) => w / k).sum() / ks.map(k => 1.0 / k).sum()
      #let ci(i, W) = calc.pow(W - ws.at(i), 2) / (2 * ks.at(i))
      #let f(W) = range(ws.len()).map(i => ci(i, W)).sum()

      #cetz.canvas({
      import cetz.draw: *
      plot.plot(
      size: (7.9, 3.8),
      x-label: [common width $W$ [pt]], y-label: [cost],
      x-min: 62, x-max: 92, y-min: 0, y-max: 0.16,
      x-tick-step: 10, y-tick-step: 0.05, y-minor-tick-step: none,
      legend: "inner-north-west", legend-style: (stroke: none),
      {
          for i in range(ws.len()) {
              plot.add(domain: (62, 92), samples: 60, style: (stroke: gray + 0.4pt),
              label: if i == 0 { $C^star_i$ } else { none }, W => ci(i, W))
              plot.add(((ws.at(i), 0),), mark: "o", mark-size: 0.1,
              mark-style: (fill: gray, stroke: none))
            }
          plot.add(domain: (62, 92), samples: 80, style: (stroke: black + 1pt), label: $F$, f)
          plot.add(((wstar, 0), (wstar, f(wstar))),
          style: (stroke: (paint: red, dash: "dashed", thickness: 0.6pt)))
          plot.add(((wstar, f(wstar)),), mark: "o", mark-size: 0.14,
          mark-style: (fill: red, stroke: none), style: (stroke: none), label: $W^star$)
        },
      )
      })
    \end{typst}
    \caption{Choosing the Common Width for Five Names.
      Each name contributes a parabola $C^\star_i$ whose vertex sits at its
      natural width $w_i$ (grey dots) and whose curvature is $1/K_i$, so stiff
      names pull weakly.
      The total $F$ is minimised at $W^\star$ (red dot), nearer the short names.}
    \label{fig:width}
  \end{figure}

  So the optimal target is neither the widest name, which would stretch everyone else, nor the arithmetic mean width, but a mean that gives most weight to the names that are most expensive to move.
  Figure~\ref{fig:width} shows the mechanism.
  Since $K_i$ grows roughly like $w_i^2$, long names are stiff, contribute little pull, and $W^\star$ is biased towards the short names - correctly, because a short name pays the largest \emph{relative} distortion for a given absolute change.

  \subsection{Feasibility and Clamping}
  Unbounded $s_i$ and $t_i$ are typographically unacceptable, so we impose $s_i \in [s_{\min}, s_{\max}]$, with defaults $0.85$ and $1.15$, and, implicitly, $t_i \ge -e_i$, at which tracking consumes the advance of an average letter.
  Letters touch well before that bound, which Section~\ref{sec:results} returns to.
  The constrained problem is still convex with a box constraint, and its solution is the projection of the unconstrained one followed by re-solving the equality constraint for the remaining free variable:
  \begin{equation}
    \hat s_i = \operatorname{clamp}(s_i, s_{\min}, s_{\max}), \qquad \hat t_i = \frac{W^\star - \hat s_i w_i}{g_i}
    \label{eq:clamp}
  \end{equation}
  for $g_i \ge 1$.
  This projection is exact.
  Along the width constraint $t_i$ is determined by $s_i$, and the cost becomes a convex quadratic in $s_i$ alone, centred at the unconstrained optimum, so clamping $s_i$ to the box minimises it.
  Single-character names ($g_i = 0$) have no gaps to track and must take $s_i = W^\star / w_i$ regardless of the bounds; the model degenerates gracefully and stays feasible.

  \section{Implementation}
  \label{sec:impl}
  \subsection{\LaTeX}
  \label{sec:latex}
  The package \code|equalauthors.sty| implements \eqref{eq:stiffness} to \eqref{eq:clamp} in \code|expl3| floating point.
  Names are collected with \tex|\equalauthorsAddAuthor|, and \tex|\equalauthorsAuthors| measures each name and computes $W^\star$ by \eqref{eq:wstar} and the split by \eqref{eq:split} and \eqref{eq:clamp}.
  Each name is then set as \tex|\scalebox{s}[1]{\textls[l]{name}}| with $s = s_i$ and $l = \ell_i = 1000\,t_i/(s_i\,\mathrm{em})$, the tracking in microtype's units\footnote{The $\mathrm{em}$ is the length named after the letter M, which in metal type was cast on a square body and so measures the nominal type size of the current font.}, where the division by $s_i$ compensates for \tex|\scalebox| scaling the inserted kerns too, and the boxes are overlapped in one box of width $W^\star$.

  Splitting a name into characters would break UTF-8 input on 8-bit engines, strip accents written as commands, such as \tex|\"o|, of their argument, and lose kerning pairs, so the package never takes a name apart and recovers the two quantities the model needs by measurement,
  \begin{equation}
    w_i = \omega_i(1, 0), \qquad g_i = \frac{\partial \omega_i}{\partial t_i} = \frac{\omega_i(1, \mathrm{em}) - \omega_i(1, 0)}{\mathrm{em}} .
    \label{eq:measure}
  \end{equation}
  The second measurement sets the name with a tracking of one em, and since \eqref{eq:width} is affine in $t_i$, two \tex|\hbox_set:Nn| measurements per name suffice.
  Microtype also places half the tracking outside the first and last letter, which would be invisible padding, so the package cancels it with a negative kern at either end.
  The remaining quotient equals the gap count, except that microtype stretches an interword space by $4/3$ of the tracking rather than counting it as two gaps, so $g_i$ comes out fractional for names that contain spaces.
  Nothing in Section~\ref{sec:solution} uses integrality, so the solution applies unchanged.
  The companion command \tex|\equalauthorsAuthorsStacked| prints the names one per line at the common width (Figure~\ref{fig:stacked}).

  Microtype rounds tracking to whole thousandths of an em, so the package re-solves the scale from $s_i (w_i + g_i \tau_i) = W^\star$, with $\tau_i$ the quantised tracking handed to microtype.
  The width thus stays exactly $W^\star$ while $s_i$ moves by about $10^{-3}$, and at a bound $\ell_i$ is rounded towards the inside.
  Weights and bounds are set with
  \begin{lstlisting}[style=latex, gobble=4]
    \equalauthorsSetup{alpha=1, beta=1, smin=0.85, smax=1.15}
  \end{lstlisting}

  \subsection{Citations}
  A reference list ranks authors exactly as a title page does, so the standard is only self-consistent if citations are superimposed as well.
  The command \tex|\equalauthorsList| takes a BibTeX-style name list \code|A and B and C|, splits it on \code|and|, and applies \eqref{eq:wstar} and \eqref{eq:split} to the result.
  The macro is placed in the \code|author| field of the \code|.bib| entry inside an extra pair of braces,
  \begin{lstlisting}[style=latex, gobble=4]
    author = {{\equalauthorsList{Erik D. Demaine and Martin L. Demaine}}}
  \end{lstlisting}
  which hides the \code|and| separators from biber and makes it pass the field through verbatim.
  Every reference of this paper is typeset this way, each with a common width $W^\star$ over its own names alone.

  \subsection{Typst}
  The Typst package provides the same functions, using \code|measure| for $w_i$, \code|scale| for $s_i$ and the \code|tracking| parameter of \code|text| for $t_i$.
  Grapheme clusters are a native notion there, so the encoding difficulties of Section~\ref{sec:latex} do not arise and $g_i$ is counted directly.
  Typst, too, applies tracking inside the scaled coordinate frame, so the tracking handed to the engine is $t_i / s_i$; the realised width is then $s_i w_i + g_i t_i$ as in \eqref{eq:width}.
  A self-check compiles the distorted names and asserts that every measured width equals $W^\star$ to within $10^{-4}$\,pt.

  Bibliographies need a different hook there.
  Typst renders references through a CSL style and exposes no element for the author list of an entry, so there is nothing to attach a \code|show| rule to.
  The hook is therefore moved into the citation style itself: \code|equalauthors.csl| is an ordinary numeric style whose only peculiarity is that it fences each entry's author list between two invisible \textsc{u+2062} characters, and the accompanying \code|equal-references| show rule matches exactly those runs and replaces them with the superimposed list.

  \subsection{Formal Verification}
  \label{sec:lean}
  Every mathematical claim of Sections~\ref{sec:model} to~\ref{sec:impl} and Appendix~\ref{sec:ablation} is machine-checked in Lean~4~\cite{demoura2021lean4} against Mathlib~\cite{mathlib2020}, in \code|lean/EqualAuthors.lean| of the accompanying repository\footnote{\repolink}, with every proof complete.
  This covers both propositions with uniqueness, the corollary, their composition into the full problem \eqref{eq:problem}, the clamp \eqref{eq:clamp}, the measurement and rounding identities of the implementations, and the containment claim of the introduction; each theorem is named after the label it proves, such as \code|prop_split|.
  The proofs need no Lagrange multipliers, since along the constraint the cost exceeds its optimum by $\frac{\alpha}{2}(s - s^\star)^2 + \frac{\beta}{2}\bigl((t - t^\star)/e\bigr)^2$, and $F(W)$ exceeds $F(W^\star)$ by $\frac12 \sum_i K_i^{-1} (W - W^\star)^2$.

  \section{Results}
  \label{sec:results}
  The author block of this paper is an equal-width superimposition of its three authors.
  Table~\ref{tab:results} reports the solution for a six-name example that adds the originators of the standard and one long name, and Figure~\ref{fig:stacked} unstacks the same six names so that the common width can be checked by eye.

  \equalauthorsSetup{alpha=1, beta=4, smin=0.5, smax=1.5}
  \equalauthorsClear
  \equalAuthorsTest

  \begin{table}[H]
    \centering
    \begin{tabular}{lrrr}
      \toprule
      Name $i$          & $w_i$ [pt]                                                 & $s_i$ & $t_i$ [pt] \\
      \midrule
      Peter Waldert     & \equalauthorsSolution{Peter Waldert}                                            \\
      Benedikt Kantz    & \equalauthorsSolution{Benedikt Kantz}                                           \\
      Thomas K\"ohler   & \equalauthorsSolution{Thomas K\"ohler}                                          \\
      Erik D. Demaine   & \equalauthorsSolution{Erik D. Demaine}                                          \\
      Martin L. Demaine & \equalauthorsSolution{Martin L. Demaine}                                        \\
      Jes\'us S. M. A.  & \equalauthorsSolution{Jes\'us Salvador Mart\'inez Alcaide}                      \\
      \bottomrule
    \end{tabular}
    \caption{Natural Widths and Optimal Distortions at $W^\star = \equalauthorsWidth$\,pt.
      The weights are $\alpha = 1$ and $\beta = 4$, and $s_i \in [0.5, 1.5]$.
      Names shorter than $W^\star$ receive $s_i > 1$ and $t_i > 0$; longer names receive both below their neutral values.}
    \label{tab:results}
  \end{table}

  \begin{figure}[H]
    \centering
    \equalauthorsAuthorsStacked
    \hspace{2em}
    \equalauthorsClear
    \equalAuthorsBase
    \equalauthorsAddAuthor{Yu Liu}
    \equalauthorsAuthorsStacked
    \caption{Stacked Names at the Common Width.
      Left, the six names of Table~\ref{tab:results}, distorted and printed separately; right, the same set with Yu Liu in place of the long name.
      All lines of a block have the common width $W^\star$ of that block.}
    \label{fig:stacked}
  \end{figure}
  \equalauthorsSetup{alpha=1, beta=1, smin=0.85, smax=1.15}

  The long name, Jes\'us Salvador Mart\'inez Alcaide, is on average twice as wide as the other five, yet at the settings of Table~\ref{tab:results} it moves $W^\star$ only from $69.8$ to $73.1$\,pt, where the arithmetic mean would move by $12$\,pt.
  Its stiffness grows like $w_i^2$, so it pulls with about a quarter of the weight of each of the others.

  The same name shows where the default bounds end.
  At $\alpha = \beta = 1$ and $s_i \in [0.85, 1.15]$, its unconstrained scale falls below $s_{\min}$, so the clamp \eqref{eq:clamp} is active and tracking absorbs the remaining $48$\,pt, about $-0.35\,e_i$ per gap.
  That is far above the bound $-e_i$, yet adjacent letters already run into one another and the name becomes illegible.
  Table~\ref{tab:results} therefore widens the scale range to $[0.5, 1.5]$ and penalises tracking four times as heavily, $\beta = 4$, which moves most of every correction onto the glyphs.
  The long name is then set at $s_i \approx 0.61$ with $t_i \approx -0.1\,e_i$, condensed but legible, and no clamp is active.

  A very short name stresses the opposite end.
  Put in place of the long name, Yu Liu, at $w_i = 30.1$\,pt, is the least stiff name of the set and pulls $W^\star$ from $69.8$ down to $48.8$\,pt, as the right block of Figure~\ref{fig:stacked} shows at the same settings.
  The other five names are condensed to between $s_i \approx 0.65$ and $0.83$, while Yu Liu reaches the bound $s_{\max} = 1.5$ and tracking adds the remainder, about $0.15\,e_i$ per gap.
  Loose tracking is the milder distortion, so the short name stays legible, although it now sets everyone else's width.

  \section{Limitations}
  The cost \eqref{eq:cost} models ugliness; it does not measure it.
  It is symmetric in the sign of the distortion, whereas readers tolerate loose tracking better than tight tracking, and condensed glyphs better than expanded ones; asymmetric weights $\alpha^\pm,\beta^\pm$ would leave the convexity argument unchanged.
  Vertical extent is likewise creditable and left untouched, since names with ascenders and descenders still claim more ink.

  We also skip opacity, which the original construction uses.
  \citeauthor{demaine2023every}~\cite{demaine2023every} set every name in semi-transparent ink, at opacity $2/3$, so that the superimposed names do not merge into a solid black block, and overlapping strokes appear darker, which keeps the individual names legible.
  We print all names at full opacity.
  Combining the two is promising future work, in particular an opacity that decreases with the number of authors, as \citeauthor{demaine2023every} suggest, since at fixed opacity the block grows darker with every name that is added.

  Finally, the standard equalises typography, not contribution; nothing here prevents a single coauthor from still doing all the work.

  \section{Conclusion}
  Superimposing author names removes order but not width.
  Requiring equal width turns the layout into a small convex program whose solution is fully explicit.
  The split between glyph scaling and tracking is the ratio $\alpha : \beta$ of the two ugliness weights (Proposition~\ref{prop:split}), and the common width is the stiffness-weighted mean \eqref{eq:wstar} of the natural widths (Proposition~\ref{prop:width}).
  Both are one line of arithmetic per name, so every author can be a first author of exactly the same size at no measurable cost in compile time.

  \printbibliography

  \appendix
  \section{Ablation Studies}
  \label{sec:ablation}
  \subsection{Scaling Only}
  Changing only the glyph scale $s$, that is, setting $t = 0$ for every name, simplifies the model at the cost of ugliness.
  The realised width of a name reduces to $\omega(s) = s w$ and its cost \eqref{eq:cost} to its first term,
  \[
    C(s) = \frac{\alpha}{2} (s-1)^2,
  \]
  by which the Lagrangian simplifies to
  \[
    \mathcal{L} = \frac{\alpha}{2} x^2 - \lambda (w x - d).
  \]
  Stationarity gives $x = \lambda w / \alpha$, which, substituted into the constraint, yields $d = w x = \lambda w^2 / \alpha$ and the optimal cost
  \[
    C^\star(W) = \frac{\alpha}{2} x^2 = \frac{(W-w)^2}{2 w^2 / \alpha},
  \]
  so $K = w^2/\alpha$, which is \eqref{eq:stiffness} without its tracking term.
  The common width then follows from \eqref{eq:wstar} as in the general case, and every name takes $s = W^\star / w$.
  The bounds on $s$ cannot be enforced here, since no other variable is left to absorb the remainder.
  Figure~\ref{fig:stacked:fixed_t} shows the result.
  Each name reaches $W^\star$ through its glyph widths alone, so letterforms differ visibly between the lines, from $s \approx 1.19$ for the shortest name to $s \approx 0.51$ for the longest.

  \equalauthorsSetup{alpha=1, beta=1, smin=0.85, smax=1.15, fix_t=true}
  \equalauthorsClear
  \equalAuthorsTest

  \begin{figure}[H]
    \centering
    \equalauthorsAuthorsStacked
    \caption{The Six Names of Table~\ref{tab:results} With Scaling Only, $t = 0$.
      All six lines have width $W^\star$.}
    \label{fig:stacked:fixed_t}
  \end{figure}

  \subsection{Tracking Only}
  Changing only the tracking $t$, that is, setting $s = 1$ for every name, simplifies the model in the same way and is uglier still.
  The realised width of a name reduces to $\omega(t) = w + g t$ and its cost \eqref{eq:cost} to its second term,
  \[
    C(t) = \frac{\beta}{2} \Bigl(\frac{t}{e}\Bigr)^2,
  \]
  by which the Lagrangian simplifies to
  \[
    \mathcal{L} = \frac{\beta}{2} \frac{t^2}{e^2} - \lambda (g t - d).
  \]
  Stationarity gives $\beta t / e^2 = \lambda g$, which, substituted into the constraint, yields $d = g t = \lambda g^2 e^2 / \beta$ and the optimal cost
  \[
    C^\star(W) = \frac{\beta}{2}\Bigl(\frac{t}{e}\Bigr)^2 = \frac{(W-w)^2}{2 g^2 e^2 / \beta},
  \]
  so $K = g^2 e^2 / \beta$, which is \eqref{eq:stiffness} without its scaling term.
  The common width is, again, computed by \eqref{eq:wstar}.
  Figure~\ref{fig:stacked:fixed_s} shows that tracking alone cannot absorb large width changes.
  The long name needs about $-0.5\,e$ per gap and collapses, while the shortest name is visibly letterspaced.

  \equalauthorsSetup{alpha=1, beta=1, smin=0.85, smax=1.15, fix_t=false, fix_s=true}
  \equalauthorsClear
  \equalAuthorsTest

  \begin{figure}[H]
    \centering
    \equalauthorsAuthorsStacked
    \caption{The Six Names of Table~\ref{tab:results} With Tracking Only, $s = 1$.
      All six lines have width $W^\star$.}
    \label{fig:stacked:fixed_s}
  \end{figure}
\end{document}